\documentclass[letter]{aa}
\usepackage[utf8]{inputenc}

\usepackage[]{natbib}
\bibpunct{(}{)}{;}{a}{}{,} 
\begin{document}

\titlerunning{Homing in on Polaris}
\title{Homing in on Polaris: A 7\,M$_\odot$ first-overtone Cepheid\\ 
entering the instability strip for the first time} 
\author{Richard I. Anderson\thanks{ESO fellow, \email{randerso@eso.org}}}
 \institute{European Southern Observatory, Karl-Schwarzschild-Str. 2, D-85748 Garching b. M\"unchen, Germany}
\date{Received 3 January 2018 / Accepted 8 March 2018}

\abstract{
    A recently presented {\it HST/FGS} parallax measurement of the Polaris system has been interpreted as evidence for the Cepheid Polaris Aa to be pulsating in the second overtone. An age discrepancy between components A and B has been noted and discussed in terms of a stellar merger.
    Here I show that the new parallax of Polaris is consistent with a simpler interpretation of Polaris as a $7\,\rm{M}_\odot,$ first-overtone, classical Cepheid near the hot boundary of the first instability strip crossing. This picture is anchored to rates of period change, the period-luminosity relation, the location in color-magnitude space, the interferometrically determined radius, spectroscopic N/C and N/O enhancements, and a dynamical mass measurement. The detailed agreement between models and data corroborates the physical association between the Cepheid and its visual companion as well as the accuracy of the {\it HST} parallax. The age discrepancy between components A and B is confirmed and requires further analysis, for example to investigate the possibility of stellar mergers in an evaporating birth cluster of which the Polaris triple system would be the remaining core.}


\maketitle

\section{Introduction}

\object{Polaris} is the closest ``classical'' Cepheid to the Sun. Given the importance of Cepheids for both stellar astrophysics and cosmology, it is unsurprising that Polaris\footnote{``Polaris'' here denotes the Cepheid component, Polaris Aa, unless explicitly stated otherwise.} has been studied extensively for more than a century. Polaris features several peculiarities, such as a very low and changing light amplitude, and abrupt variations in its fast changing period. A particular point of contention in the literature has been its pulsation mode and parallax, since these two are intricately linked via the period-luminosity relation (PLR). 

Polaris has at least two companions that are discussed in the literature; see \citet{1996JRASC..90..140K} for a review. One is a visual companion at a separation of approximately $18"$ \citep[Polaris B]{1966AJ.....71..732F,2008AJ....136.1137E}, and the other is a much more nearby companion that was first discovered spectroscopically \citep{1929PASP...41...56M} and later spatially resolved approximately $0.17"$ from the Cepheid \citep[Polaris Ab]{2008AJ....136.1137E}. Although the separations are numerically different, the Polaris system thus shares an important similarity with the prototype \object{$\delta$ Cephei} \citep{2015ApJ...804..144A}. 

Recently, \citet[henceforth:B18]{2018ApJ...853...55B} provided an independent parallax measurement of \object{Polaris B}, the visual companion generally thought to be physically associated with the Cepheid. Intriguingly, this new parallax is much smaller than the {\it Hipparcos} parallax, that is, $6.26 \pm 0.24$\,mas (B18) versus $7.54 \pm 0.11$\,mas \citep{2007ASSL..350.....V}, respectively. Another parallax estimate of $10.1 \pm 0.2$\,mas \citep{2013ApJ...762L...8T} based on putative cluster membership has been vigorously disputed \citep{2013A&A...550L...3V}. 

The B18 parallax of Polaris B should be applicable for the Cepheid, if Polaris A and B are physically associated. This assumption appears warranted based on several membership indicators, such as photometry, proper motion, and radial velocities; see B18 and references therein. Given the B18 parallax, the physical separation between A and B is $\sim 2800$\,au, which is  within the range of physically associated wide binaries among Cepheids \citep[$a_{\rm{rel}} \lesssim 4000$\,au]{2016AJ....151..108E,2016AJ....151..129E}. However, this poses a problem, since Polaris B (age $> 2.1$\,Gyr) is much older than the Cepheid (B18). Moreover, reconciling the B18 and {\it Hipparcos} parallaxes seems impossible, since the two values differ by $4.8\,\sigma$. 

B18 argue that Polaris should be pulsating in the second overtone based on its absolute $V-$band magnitude and pulsation period, $P$. However, B18 also note that the 2O pulsation mode for Polaris is unlikely, since 2O pulsators tend to exhibit more than a single pulsation mode and since Polaris would have an unusually long period for a Galactic 2O pulsator. To explain the age discrepancy between components A and B, B18 discuss the possibility of mergers among system components, although no final conclusion could be reached. 

This {\it letter} is structured as follows. \S\ref{sec:models} provides period-age and period-radius relations of first overtone Cepheids that cross the classical instability strip (IS) for the first time based on Geneva stellar evolution models that incorporate the effects of rotation. \S\ref{sec:consistent} compares the observed properties of Polaris\hbox{---}adopting the B18 parallax for the Cepheid\hbox{---}with these model predictions, and draws a consistent picture of its evolutionary status. \S\ref{sec:discussion} discusses how this picture relates to a dynamical mass measurement and the age discrepancy. \S\ref{sec:conclusions} summarizes this work.

\section{Model predictions}\label{sec:models}

\citet[henceforth: A16]{2016A&A...591A...8A} recently provided a detailed linear non-adiabatic radial pulsation analysis of Geneva stellar evolution models that incorporate the effects of rotation on stellar evolution. Rotation is a key ingredient in stellar models that remains under-discussed and whose influence on the evolution of Cepheids has been shown to be very significant. For instance, Geneva stellar evolution models \citep{2012A&A...537A.146E,2013A&A...553A..24G} do not exhibit a mass discrepancy compared to other mass estimates \citep[A16]{2014A&A...564A.100A,2017EPJWC.15206002A}. 

A16 published predictions for the position of the classical IS as well as period-age, period-radius, and of course period-luminosity relations, among others, for Cepheids of different metallicities (Solar, LMC, SMC) and initial rotation rates (no, average, and fast rotation) that pulsate in the first overtone (FO) or the fundamental (FU) mode. Moreover, A16 distinguished between Cepheids crossing the IS for the first, second, and third time, and provided analytic expressions for different relations with $P$\hbox{ ; }mainly for second and third crossings as these are the most common among classical Cepheids. 

To evaluate whether the observed properties of Polaris, which is arguably \emph{not} representative of the average Cepheid, are consistent with the Geneva models, the following provides analytic expressions for period-age (\S\ref{sec:pa}) and period-radius (\S\ref{sec:pr}) relations of FO Cepheids crossing the IS for the first time. These relations have been obtained by fitting the published results from A16.

\subsection{Period-age relations}\label{sec:pa}
Table\,\ref{tab:period-age} provides analytic period-age relations for FO Cepheids on the first IS crossing with average initial rotation rate $\omega = \Omega / \Omega_{\rm{crit}} = 0.5$. 
For Polaris, assuming Solar metallicity ($Z_\odot$), average initial rotation, FO pulsation and a first IS crossing very near the hot IS boundary (cf. below), these relations yield an age of $54$\,Myr, that is, $\log{(a\,\rm{[yr]})} = 7.73$. For comparison, period-age relations based on models with no and fast ($\omega = 0.9$) rotation would imply ages of $35$ and $47$\,Myr, respectively.

\subsection{Period-radius relations}\label{sec:pr}

Table\,\ref{tab:period-radius} provides analytic period-radius relations for FO Cepheids on the first IS crossing with average initial rotation rate $\omega = \Omega / \Omega_{\rm{crit}} = 0.5$. 
For Polaris, assuming $Z_\odot$, average initial rotation, FO pulsation and a first IS crossing very near the hot IS boundary (cf. below), these relations yield a radius of $50.6\,\rm{R}_\odot$. We note, however, that the fitted linear relation predicts a slightly smaller radius than the closest computed model; see Sec.\,\ref{sec:radius}. Period-radius relations for $\omega = 0.0$ and $0.9$ predict values of $53.8$ and $53.2\,$R$_\odot$.

\section{A consistent picture for a peculiar Cepheid}\label{sec:consistent}
This section compares observed properties of Polaris with predictions from Geneva stellar evolution models adopting the B18 parallax for the Cepheid Polaris.

\subsection{Polaris and the first IS crossing}\label{sec:pdot}

Rates of changing pulsation periods are commonly interpreted as evidence for secular evolution, i.e., the large-scale evolution of a Cepheid's radius as it passes through the IS. Although Cepheids exhibit a diverse phenomenology of observed period changes on short time scales (years to a decade) \citep[e.g.,][]{2008AcA....58..313P,2017arXiv170310334S}, stellar evolution models generally provide a good match to rates of period change determined using temporal baselines of multiple decades (e.g., A16). Positive $\dot{P}$ implies a rightward motion in the Hertzsprung-Russell diagram, that is, an increase in stellar radius. Since the first IS crossing has a particularly short lifetime, $\dot{P}$ is expected to be $1.5 - 2$ orders of magnitude faster on the first crossing than on the third.

As noted previously in the literature, the rate at which Polaris' period changes is comparatively fast, on the order of $4.4$ to $4.9\,\rm{s\,yr}^{-1}$ (based on data spanning $\sim 150$\,years) \citep[e.g.,][and references therein]{2002ApJ...567.1121E,2005PASP..117..207T,2008MNRAS.388.1239S,2008ApJ...683..433B}. Of course, Polaris has previously been discussed as a Cepheid on the first IS crossing \citep[e.g.,][A16]{2013ApJ...762L...8T,2015MNRAS.449.1011F}. A peculiar aspect of Polaris' period change is the observed discontinuity in the observed O-C diagram parabola, which implies a sudden change in pulsation period to have occurred around 1963. Rapid monitoring using the {\it SMEI} instrument on board the {\it Coriolis} spacecraft suggests that another such break may have occurred more recently \citep{2008MNRAS.388.1239S,2008ApJ...683..433B}. Polaris is furthermore famous for its changing light amplitude, which for a while was thought to be disappearing. 

\begin{figure}
    \centering
    \includegraphics{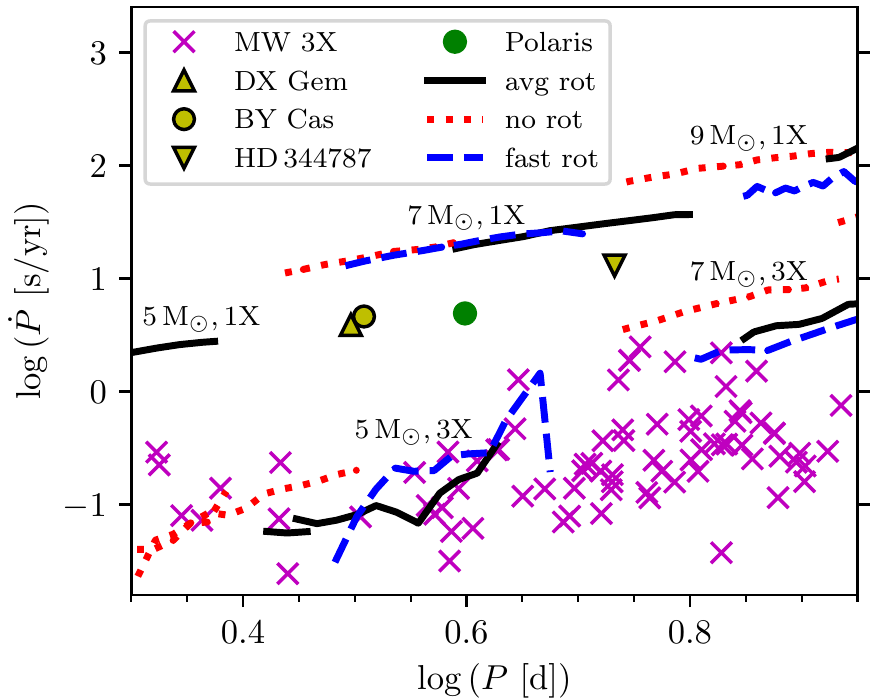}
    \caption{Predicted rates of period change for $Z_\odot$ FO Cepheids (lines) and observed values for Galactic Cepheids \citep[markers include both FO and FU Cepheids]{2006PASP..118..410T}. Adapted from Figure 13 in A16 to illustrate predictions for FO Cepheids. First IS crossings are labeled `1X', third IS crossings `3X'. Polaris (green filled circle) and three other Cepheids (\object{DX Gem}, \object{BY Cas}, and \object{HD 344787}; yellow markers) exhibit much faster $\dot{P}$ than the bulk of Cepheids and are likely crossing the IS for the first time. Line styles distinguish initial rotation rates of the models: dotted red = no rotation, solid black = average rotation ($\omega = 0.5$), dashed blue = fast rotation ($\omega = 0.9$). Groups of lines correspond to different initial mass or IS crossing.
    Each line segment represents an IS crossing (left is hot IS boundary). 
    }
    \label{fig:Pdot}
\end{figure}

Figure\,\ref{fig:Pdot} shows model predictions for $Z_\odot$ FO Cepheids of different initial (ZAMS) rotation rates $\omega$ on the first (faster $\dot{P}$) and third IS crossings. Predicted values of $\dot{P}$ for FU Cepheids are very similar, if slightly lower and are not shown here (cf. A16). The observed $\dot{P}$ values of four Cepheids, one of which is Polaris, by far exceed the bulk of objects consisting of both FO and FU Cepheids. The predicted $\dot{P}$ for the first crossing $7\,\rm{M}_\odot$ model is considerably faster than the observed value, and a similar average tendency is seen among third-crossing models. Hence, it appears that models systematically overestimate $\dot{P}$, that is, they underestimate evolutionary timescales. The same effect is seen also for other models \citep[cf. A16]{2013AstL...39..746F,2014AstL...40..301F}. Despite the mismatch in absolute terms, the clear separation into two groups likely indicates that Polaris is crossing the IS for the first time.

\subsection{Polaris as a first overtone pulsator}\label{sec:mode}

\begin{figure}
    \centering
    \includegraphics{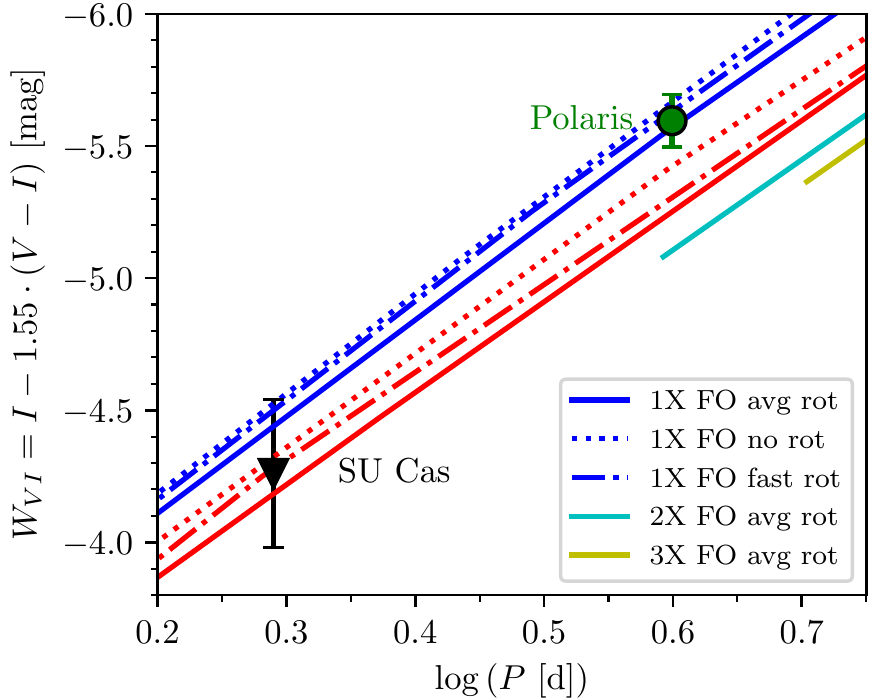}
    \caption{The position of Polaris on the predicted reddening-free Period-Wesenheit relation for FO Cepheids on the first crossing; relations for models on the blue and red IS edges are shown in these colors. Solid lines are model predictions featuring average rotation (half of critical angular rotation rate on ZAMS). Models without rotation or very fast rotation are shown by dotted and dash-dotted lines and are very similar near the blue edge. SU\,Cas is another FO Cepheid with positive $\dot{P}$ (likely on second crossing) and available parallax \citep[{\it Hipparcos}]{2007ASSL..350.....V}. PLRs near red IS boundaries for second and third crossing FO Cepheids are shown in cyan and yellow. Blue boundaries for second or third crossings could not be established at such short periods, since their blue loops are not extended to sufficiently high temperatures \citep[cf.][]{2017EPJWC.15206002A}.}
    \label{fig:FO-PLR}
\end{figure}

The pulsation mode of Polaris has been a matter of intense discussion, in particular with regards to the calibration of the PLR. Adopting the {\it HST} parallax, B18 noted the difficulty of rendering the absolute magnitude of Polaris consistent with the PLR of either FU or FO Cepheids, while implicitly assuming that Polaris crosses the IS for the third time. 
Figure\,\ref{fig:FO-PLR} shows the position of Polaris in the period-luminosity relation using the optical reddening-free Wesenheit magnitude $W_{\rm{VI}} = I - 1.55\cdot(V-I)$ \citep{1982ApJ...253..575M,2008AcA....58..163S}. As the figure shows, the position of Polaris agrees with the predictions for a FO Cepheid on the first IS crossing very near the hot IS boundary. The much shorter-period FO Cepheid SU\,Cas (on the second IS crossing) lies below the period range where $Z_\odot$ models exhibit blue loops sufficiently extended for predicting blue IS boundaries; see \citet{2017EPJWC.15206002A} for a discussion.

\begin{figure*}
\centering
\includegraphics{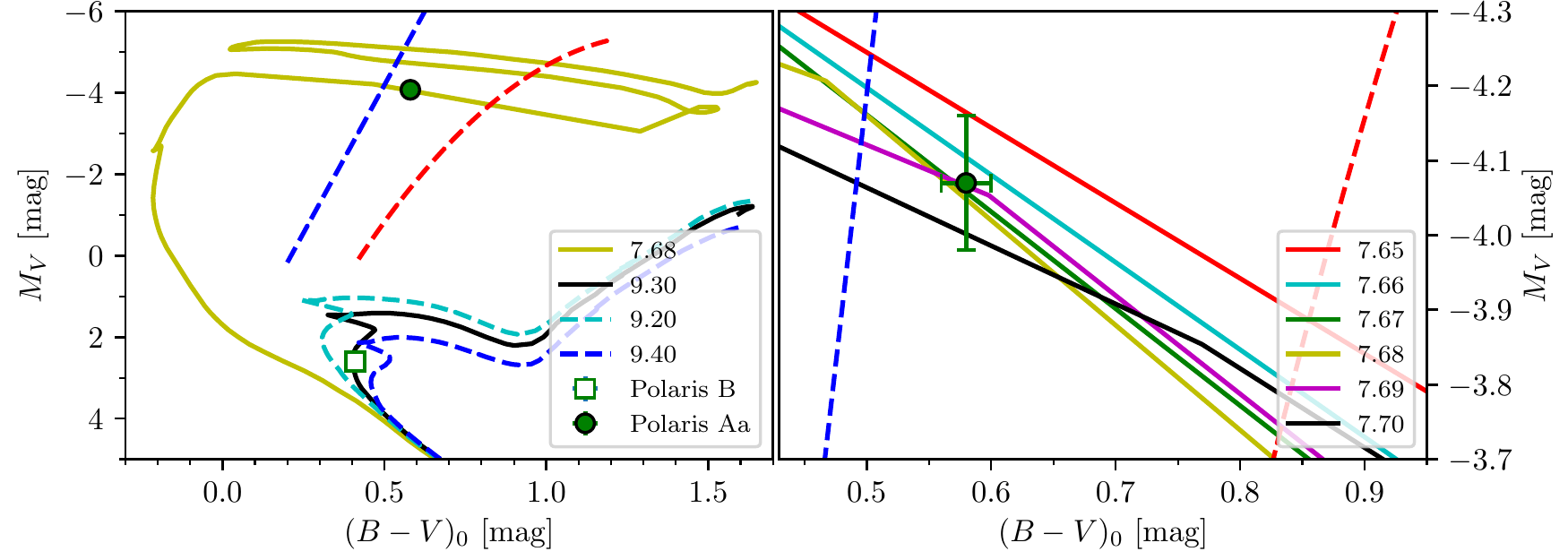}
\caption{Color-magnitude diagram for Polaris with predicted $Z_\odot$ isochrones of different $\log{a}$ as labeled. The position of Polaris B is clearly discrepant with the isochrone that matches Polaris Aa, indicating a significant difference in age  (best fit isochrone for $\log{a} = 9.3$). $M_V$ and $(B-V)_0$ values for Polaris Aa and B taken from B18.}
\label{fig:CMD}
\end{figure*}

Figure\,\ref{fig:CMD} shows isochrones computed using the Geneva stellar isochrone online interpolation tool\footnote{\url{https://obswww.unige.ch/Recherche/evoldb/index/}} \citep{2012A&A...537A.146E,2014A&A...566A..21G}  assuming average initial rotation rates together with the $M_V$ and $(B-V)_0$ values from B18. These isochrones were computed for a range of ages near $\log{a} = 7.73$; see \S\ref{sec:pa} and Table\,\ref{tab:period-age}. The right panel shows a closeup of the isochrones that pass through the uncertainties of Polaris Aa. Instability strip boundaries\footnote{FO Cepheid on first crossing hot boundary: $M_V = 3.074 - 14.522 \cdot (B-V)$; cool boundary: $M_V = 6.247 - 17.376 \cdot (B-V) + 6.464 \cdot (B-V)^2$.} were determined by fitting predictions for all rotation rates of FO Cepheids on the first crossing based on Geneva evolution models (A16), and are thus self-consistent with the isochrones. 
The isochrone suggests a slightly younger age of $\log{a} \in [7.65, 7.70]$ than the $P-a$ relation. As noted by B18, the position of Polaris B in the CMD does not match the isochrone for the Cepheid, regardless of the assumed age or IS crossing.

\subsection{The radius of Polaris}\label{sec:radius}
\begin{figure}
    \centering
    \includegraphics{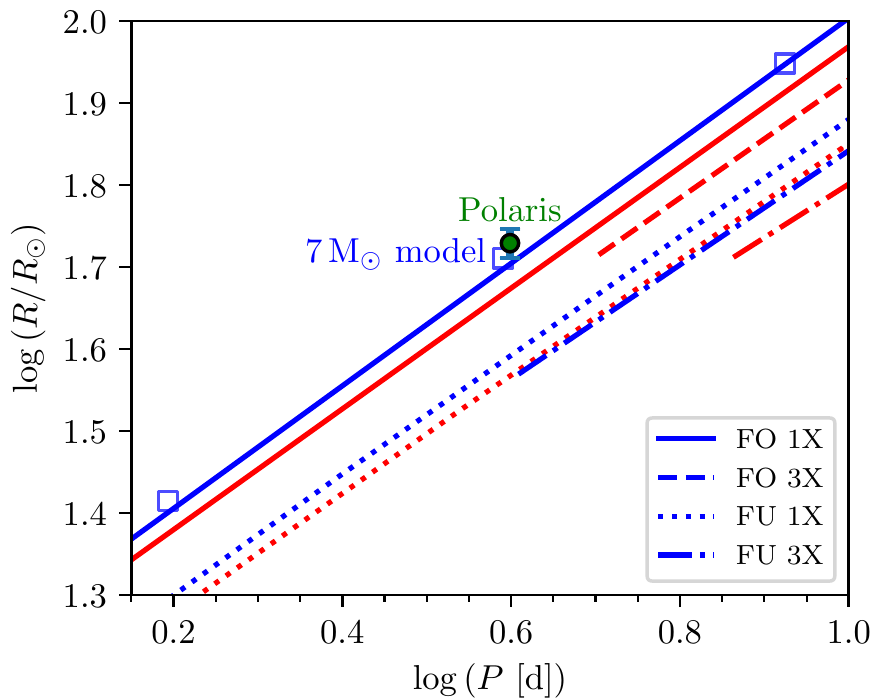}
    \caption{Predicted $Z_\odot$ period-radius relations and Polaris' interferometrically measured radius assuming the B18 parallax. Blue and red lines drawn are linear fits to the discretely computed models near the hot and cool IS edges; blue open squares mark computed models along the hot boundary of the first crossing. 1X denotes first crossing models, 3X third crossing models. $P-R$ relations for FU Cepheids are shown for reference. The period and radius of a $7\,\rm{M}_\odot$ 1X model ($\log{P} = 0.5904$, $\log{R/R_\odot} = 1.7107$) on the first crossing nearly coincides with the Polaris measurement ($0.5987$, $1.7114$).}
    \label{fig:P-R}
\end{figure}

The fitted period-radius relation for $Z_\odot$ in Table \ref{tab:period-radius} predicts $R = 50.6\,\rm{R}_\odot$ for $P= 3.969$\,d  near the hot IS edge.
However, the predicted radius of a computed $7\,\rm{M}_\odot$ FO model (with average initial rotation) at the first crossing's hot IS boundary is $R = 51.4\,\rm{R}_\odot$ (Fig.\,\ref{fig:P-R}), and matches the interferometrically measured radius assuming the B18 parallax to within the uncertainty \citep[$\theta_{\rm{LD}} = 3.123 \pm 0.008$\,mas yields $R_{\rm{obs}} = 53.6 \pm 2.2\,\rm{R}\odot$]{2006A&A...453..155M}. The period of the same model and the Cepheid differ by less than $2\%$ (predicted $P=3.895$\,d). Period-radius relations for no or very fast rotation match the observed radius even better (see Sec.\,\ref{sec:pr}) and would imply higher and lower mass, respectively. However, the observed rotational enhancement of N/C and N/O (Sec.\,\ref{sec:mixing}) strongly favors a typical rate of initial angular rotation and the (computed) $7\,$M$_\odot$ solution.

\subsection{N/C and N/O enhancement due to rotational mixing}\label{sec:mixing}
Pre-dredge-up enhancement of nitrogen relative to carbon and oxygen provides ``smoking gun'' evidence for the presence of rotational mixing, since N has the slowest destruction rate in the CNO cycle \citep{1985ESOC...21..187M,2000ARA&A..38..143M}. Estimating this enhancement using measured CNO element abundances \citep{2005MNRAS.362.1219U} and assuming a Solar ZAMS composition (since [Fe/H] = 0.07) yields $\Delta \rm{[N/C]} = 0.59 \pm 0.16$ and $\Delta \rm{[N/O]} = 0.42 \pm 0.15$. These numbers are in excellent agreement  with the rotational enhancement predicted by the $\log{a}=7.68$ isochrone ($\Delta \rm{[N/C]}= 0.54$, $\Delta \rm{[N/O]}= 0.42$)\footnote{For an illustration of the dependence of these numbers on $\omega$ for a $7\,\rm{M}_\odot$ $Z_\odot$ Cepheid, see \citet[Fig.\,8]{2014A&A...564A.100A}.} near the first crossing of the instability strip, thus contradicting previous results \citep{2014A&A...563A..48N}. \S\ref{sec:radius} and \S\ref{sec:pdot} suggested the same value for mass. 
For comparison, an isochrone for a $47$\,Myr star entering the IS for the first time computed using fast rotation ($\omega=0.9$) predicts $\Delta\rm{[N/C]}=1.17$ and $\Delta \rm{[N/O]}=0.75$, far higher than the spectroscopically measured values.

\section{Discussion}\label{sec:discussion}

\subsection{Mass of the Cepheid Polaris Aa}\label{sec:mass}

The mass of Polaris implied by this work is very near $7\,\rm{M}_\odot$. This is slightly above the $1\sigma$ range of the  $4.5^{+2.2}_{-1.4}\,\rm{M}_\odot$ literature value measured using a spectroscopic orbit, proper motions, and assuming $\varpi = 7.72\pm0.12$\,mas \citep[E08]{2008AJ....136.1137E}. Adopting the B18 parallax increases the total mass of the Polaris system by nearly a factor of two, that is, from $5.8$ to $10.9\,\rm{M}_\odot$ (E08, Joint Fit) or from $3.9$ to $7.3\,\rm{M}_\odot$ (E08, {\it HST} Only). Since the B18 parallax also implies a greater luminosity for Polaris B, its previous mass estimate of $1.35\,\rm{M}_\odot$ also requires upward correction. Considering the (large) uncertainties of Polaris' mass measurement and its need for upward revision, the predicted mass appears to be in very good agreement with observations. 

\subsection{The age discrepancy between Polaris Aa and B}\label{sec:capture}

B18 noted an age difference between Polaris Aa ($\sim 50\,$Myr) and B ($\sim 2.1$\,Gyr) and discussed different possibilities of explaining this discrepancy via stellar mergers. The present work (Fig.\,\ref{fig:CMD}) confirms that age discrepancy using Geneva stellar evolution models.

The Polaris system consists of a $\sim 7\,\rm{M}_\odot$ Cepheid (this work) and two $\sim 1.3\,\rm{M}_\odot$ companions (E08): Polaris Ab (F6V) and Polaris B (F3V). Given the matching spectral types and similar masses of Polaris Ab and B, it seems most likely that the Cepheid would be the product of a merger.  However, a merger scenario involving two stars of   $\sim 3-4\,\rm{M}_\odot$
 does not resolve the age discrepancy, since such stars have a lifetime of only about $500$\,Myr, leaving the increasingly unlikely scenario of multiple mergers.

Interpreting the Polaris system ($M_{\rm{tot}} \approx 11\,\rm{M}_\odot$, cf. \S\ref{sec:mass}) as the remaining core of a dispersed birth cluster would imply an original cluster mass $M_{\rm{cl}} \approx 20-40\,\rm{M}_\odot$ and thus a maximum stellar mass in the order of $3-4$\,M$_\odot$ \citep[Fig.\,4-5]{2013pss5.book..115K}.
Plotting all three components (Aa, Ab, and B) in a single CMD would help establish whether component Ab is co-eval with the Cepheid or Polaris B. The high eccentricity of the inner binary (Aa-Ab) together with the presence of a far outward component B provides evidence of Kozai-Lidov interactions that could have facilitated a merger in the inner system. Stimulated evolution \citep{1995MNRAS.277.1507K} could have also been relevant for this system. The presence of a significant ($1.5 \pm 0.4\,\%$ $K-$band flux contribution) circumstellar environment around Polaris Aa \citep[located at 2.4 times the Cepheid's present radius]{2006A&A...453..155M} may further point to a merger event, which would have taken place within the last $\sim 50$\,Myr. 
Dynamical modeling of low-mass clusters would provide important insights into such systems.

In this context, two things are especially worth noting. First, there does not seem to be a need to invoke any peculiar evolutionary scenarios from the point of view of the Cepheid alone, given the above excellent agreement between observations and predictions. Thus, if a merger occurred in the past, then single star evolution models (of a discrepant age) are sufficient to explain the Cepheid's current properties. Second, there are striking similarities between the Polaris system and \object{$\delta$\,Cephei}, which is also a wide triple system with an eccentric inner binary \citep{2015ApJ...804..144A} and a nearby overdensity of dispersed young stars \citep{1999AJ....117..354D,GaiaClustersAstrometry}. Moreover, an important CSE around $\delta$\,Cephei has been detected using different techniques \citep{2006A&A...453..155M,2010ApJ...725.2392M,2012ApJ...744...53M,2016A&A...593A..45N}. 

\section{Conclusions}\label{sec:conclusions}
Model predictions for a $7\,\rm{M}_\odot$ $Z_\odot$ FO Cepheid with typical initial rotation near the hot edge of the first IS crossing provide a consistent picture for the observed properties of the enigmatic classical Cepheid Polaris when assuming that the recently measured {\it HST/FGS} parallax of its visual companion Polaris B also applies to the Cepheid. This consistent picture is anchored to rates of period change, the period-luminosity-relation, the location in color-magnitude space, the interferometrically measured radius, spectroscopic N/C and N/O enhancements, and the dynamical mass measurement from the literature. 

The strong level of agreement between models and data corroborates independent evidence supporting the physical association between Polaris A and B and the accuracy of the B18 parallax. This represents a success of stellar evolution models that are particularly sensitive to Cepheid properties. As a star that has only recently turned off from the Main Sequence, Polaris is sure to provide important constraints on rotational mixing by comparing its properties to Cepheids on later IS crossings. The age discrepancy between components Aa and B points to dynamical effects that require further study, via N-body simulations, for example. 

\begin{acknowledgements} 
The author thanks the anonymous referee for a constructive report and Nancy R. Evans for drawing his attention to the recent B18 parallax result and the age discrepancy between Polaris Aa and B. Useful discussions with Pavel Kroupa, Hideyuki Saio, Antoine M\'erand, Martino Romaniello, and Henri Boffin are acknowledged.

This research has made use of NASA's Astrophysics Data System.
\end{acknowledgements}

\bibliographystyle{aa} 
\bibliography{biblio}

\begin{thebibliography}{41}
\expandafter\ifx\csname natexlab\endcsname\relax\def\natexlab#1{#1}\fi

\bibitem[{{Anderson} {et~al.}(2014){Anderson}, {Ekstr{\"o}m}, {Georgy},
  {Meynet}, {Mowlavi}, \& {Eyer}}]{2014A&A...564A.100A}
{Anderson}, R.~I., {Ekstr{\"o}m}, S., {Georgy}, C., {et~al.} 2014, \aap, 564,
  A100

\bibitem[{{Anderson} {et~al.}(2017){Anderson}, {Ekstr{\"o}m}, {Georgy},
  {Meynet}, \& {Saio}}]{2017EPJWC.15206002A}
{Anderson}, R.~I., {Ekstr{\"o}m}, S., {Georgy}, C., {Meynet}, G., \& {Saio}, H.
  2017, in European Physical Journal Web of Conferences, Vol. 152, European
  Physical Journal Web of Conferences, 06002

\bibitem[{{Anderson} {et~al.}(2015){Anderson}, {Sahlmann}, {Holl}, {Eyer},
  {Palaversa}, {Mowlavi}, {S{\"u}veges}, \& {Roelens}}]{2015ApJ...804..144A}
{Anderson}, R.~I., {Sahlmann}, J., {Holl}, B., {et~al.} 2015, \apj, 804, 144

\bibitem[{{Anderson} {et~al.}(2016){Anderson}, {Saio}, {Ekstr{\"o}m}, {Georgy},
  \& {Meynet}}]{2016A&A...591A...8A}
{Anderson}, R.~I., {Saio}, H., {Ekstr{\"o}m}, S., {Georgy}, C., \& {Meynet}, G.
  2016, \aap, 591, A8

\bibitem[{{Bond} {et~al.}(2018){Bond}, {Nelan}, {Remage Evans}, {Schaefer}, \&
  {Harmer}}]{2018ApJ...853...55B}
{Bond}, H.~E., {Nelan}, E.~P., {Remage Evans}, N., {Schaefer}, G.~H., \&
  {Harmer}, D. 2018, \apj, 853, 55

\bibitem[{{Bruntt} {et~al.}(2008){Bruntt}, {Evans}, {Stello}, {Penny}, {Eaton},
  {Buzasi}, {Sasselov}, {Preston}, \& {Miller-Ricci}}]{2008ApJ...683..433B}
{Bruntt}, H., {Evans}, N.~R., {Stello}, D., {et~al.} 2008, \apj, 683, 433

\bibitem[{{de Zeeuw} {et~al.}(1999){de Zeeuw}, {Hoogerwerf}, {de Bruijne},
  {Brown}, \& {Blaauw}}]{1999AJ....117..354D}
{de Zeeuw}, P.~T., {Hoogerwerf}, R., {de Bruijne}, J.~H.~J., {Brown}, A.~G.~A.,
  \& {Blaauw}, A. 1999, \aj, 117, 354

\bibitem[{{Ekstr{\"o}m} {et~al.}(2012){Ekstr{\"o}m}, {Georgy}, {Eggenberger},
  {Meynet}, {Mowlavi}, {Wyttenbach}, {Granada}, {Decressin}, {Hirschi},
  {Frischknecht}, {Charbonnel}, \& {Maeder}}]{2012A&A...537A.146E}
{Ekstr{\"o}m}, S., {Georgy}, C., {Eggenberger}, P., {et~al.} 2012, \aap, 537,
  A146

\bibitem[{{Evans} {et~al.}(2016{\natexlab{a}}){Evans}, {Bond}, {Schaefer},
  {Mason}, {Tingle}, {Karovska}, \& {Pillitteri}}]{2016AJ....151..129E}
{Evans}, N.~R., {Bond}, H.~E., {Schaefer}, G.~H., {et~al.} 2016{\natexlab{a}},
  \aj, 151, 129

\bibitem[{{Evans} {et~al.}(2016{\natexlab{b}}){Evans}, {Pillitteri}, {Wolk},
  {Karovska}, {Tingle}, {Guinan}, {Engle}, {Bond}, {Schaefer}, \&
  {Mason}}]{2016AJ....151..108E}
{Evans}, N.~R., {Pillitteri}, I., {Wolk}, S., {et~al.} 2016{\natexlab{b}}, \aj,
  151, 108

\bibitem[{{Evans} {et~al.}(2002){Evans}, {Sasselov}, \&
  {Short}}]{2002ApJ...567.1121E}
{Evans}, N.~R., {Sasselov}, D.~D., \& {Short}, C.~I. 2002, \apj, 567, 1121

\bibitem[{{Evans} {et~al.}(2008){Evans}, {Schaefer}, {Bond}, {Bono},
  {Karovska}, {Nelan}, {Sasselov}, \& {Mason}}]{2008AJ....136.1137E}
{Evans}, N.~R., {Schaefer}, G.~H., {Bond}, H.~E., {et~al.} 2008, \aj, 136, 1137

\bibitem[{{Fadeyev}(2013)}]{2013AstL...39..746F}
{Fadeyev}, Y.~A. 2013, Astronomy Letters, 39, 746

\bibitem[{{Fadeyev}(2014)}]{2014AstL...40..301F}
{Fadeyev}, Y.~A. 2014, Astronomy Letters, 40, 301

\bibitem[{{Fadeyev}(2015)}]{2015MNRAS.449.1011F}
{Fadeyev}, Y.~A. 2015, \mnras, 449, 1011

\bibitem[{{Fernie}(1966)}]{1966AJ.....71..732F}
{Fernie}, J.~D. 1966, \aj, 71, 732

\bibitem[{{Gaia Collaboration} {et~al.}(2017){Gaia Collaboration}, {van
  Leeuwen}, {Vallenari}, \& et~al.}]{GaiaClustersAstrometry}
{Gaia Collaboration}, {van Leeuwen}, F., {Vallenari}, A., \& et~al. 2017, \aap,
  601, A19

\bibitem[{{Georgy} {et~al.}(2013){Georgy}, {Ekstr{\"o}m}, {Granada}, {Meynet},
  {Mowlavi}, {Eggenberger}, \& {Maeder}}]{2013A&A...553A..24G}
{Georgy}, C., {Ekstr{\"o}m}, S., {Granada}, A., {et~al.} 2013, \aap, 553, A24

\bibitem[{{Georgy} {et~al.}(2014){Georgy}, {Granada}, {Ekstr{\"o}m}, {Meynet},
  {Anderson}, {Wyttenbach}, {Eggenberger}, \& {Maeder}}]{2014A&A...566A..21G}
{Georgy}, C., {Granada}, A., {Ekstr{\"o}m}, S., {et~al.} 2014, \aap, 566, A21

\bibitem[{{Kamper}(1996)}]{1996JRASC..90..140K}
{Kamper}, K.~W. 1996, \jrasc, 90, 140

\bibitem[{{Kroupa}(1995)}]{1995MNRAS.277.1507K}
{Kroupa}, P. 1995, \mnras, 277 [\eprint{astro-ph/9508084}]

\bibitem[{{Kroupa} {et~al.}(2013){Kroupa}, {Weidner}, {Pflamm-Altenburg},
  {Thies}, {Dabringhausen}, {Marks}, \& {Maschberger}}]{2013pss5.book..115K}
{Kroupa}, P., {Weidner}, C., {Pflamm-Altenburg}, J., {et~al.} 2013, in Planets,
  Stars and Stellar Systems.~Volume 5: Galactic Structure and Stellar
  Populations, ed. T.~D. {Oswalt} \& G.~{Gilmore} (Springer, Dordrecht), 115

\bibitem[{{Madore}(1982)}]{1982ApJ...253..575M}
{Madore}, B.~F. 1982, \apj, 253, 575

\bibitem[{{Maeder}(1985)}]{1985ESOC...21..187M}
{Maeder}, A. 1985, in European Southern Observatory Conference and Workshop
  Proceedings, Vol.~21, European Southern Observatory Conference and Workshop
  Proceedings, ed. I.~J. {Danziger}, F.~{Matteucci}, \& K.~{Kjar}, 187--205

\bibitem[{{Maeder} \& {Meynet}(2000)}]{2000ARA&A..38..143M}
{Maeder}, A. \& {Meynet}, G. 2000, \araa, 38, 143

\bibitem[{{Marengo} {et~al.}(2010){Marengo}, {Evans}, {Barmby}, {Matthews},
  {Bono}, {Welch}, {Romaniello}, {Huelsman}, {Su}, \&
  {Fazio}}]{2010ApJ...725.2392M}
{Marengo}, M., {Evans}, N.~R., {Barmby}, P., {et~al.} 2010, \apj, 725, 2392

\bibitem[{{Matthews} {et~al.}(2012){Matthews}, {Marengo}, {Evans}, \&
  {Bono}}]{2012ApJ...744...53M}
{Matthews}, L.~D., {Marengo}, M., {Evans}, N.~R., \& {Bono}, G. 2012, \apj,
  744, 53

\bibitem[{{M{\'e}rand} {et~al.}(2006){M{\'e}rand}, {Kervella}, {Coud{\'e} du
  Foresto}, {Perrin}, {Ridgway}, {Aufdenberg}, {ten Brummelaar}, {McAlister},
  {Sturmann}, {Sturmann}, {Turner}, \& {Berger}}]{2006A&A...453..155M}
{M{\'e}rand}, A., {Kervella}, P., {Coud{\'e} du Foresto}, V., {et~al.} 2006,
  \aap, 453, 155

\bibitem[{{Moore}(1929)}]{1929PASP...41...56M}
{Moore}, J.~H. 1929, \pasp, 41, 56

\bibitem[{{Nardetto} {et~al.}(2016){Nardetto}, {M{\'e}rand}, {Mourard},
  {Storm}, {Gieren}, {Fouqu{\'e}}, {Gallenne}, {Graczyk}, {Kervella},
  {Neilson}, {Pietrzynski}, {Pilecki}, {Breitfelder}, {Berio}, {Challouf},
  {Clausse}, {Ligi}, {Mathias}, {Meilland}, {Perraut}, {Poretti}, {Rainer},
  {Spang}, {Stee}, {Tallon-Bosc}, \& {ten Brummelaar}}]{2016A&A...593A..45N}
{Nardetto}, N., {M{\'e}rand}, A., {Mourard}, D., {et~al.} 2016, \aap, 593, A45

\bibitem[{{Neilson}(2014)}]{2014A&A...563A..48N}
{Neilson}, H.~R. 2014, \aap, 563, A48

\bibitem[{{Poleski}(2008)}]{2008AcA....58..313P}
{Poleski}, R. 2008, \actaa, 58, 313

\bibitem[{{Soszynski} {et~al.}(2008){Soszynski}, {Poleski}, {Udalski},
  {Szymanski}, {Kubiak}, {Pietrzynski}, {Wyrzykowski}, {Szewczyk}, \&
  {Ulaczyk}}]{2008AcA....58..163S}
{Soszynski}, I., {Poleski}, R., {Udalski}, A., {et~al.} 2008, \actaa, 58, 163

\bibitem[{{Spreckley} \& {Stevens}(2008)}]{2008MNRAS.388.1239S}
{Spreckley}, S.~A. \& {Stevens}, I.~R. 2008, \mnras, 388, 1239

\bibitem[{{S{\"u}veges} \& {Anderson}(2017)}]{2017arXiv170310334S}
{S{\"u}veges}, M. \& {Anderson}, R.~I. 2017, ArXiv e-prints
  [\eprint[arXiv]{1703.10334}]

\bibitem[{{Turner} {et~al.}(2006){Turner}, {Abdel-Sabour Abdel-Latif}, \&
  {Berdnikov}}]{2006PASP..118..410T}
{Turner}, D.~G., {Abdel-Sabour Abdel-Latif}, M., \& {Berdnikov}, L.~N. 2006,
  \pasp, 118, 410

\bibitem[{{Turner} {et~al.}(2013){Turner}, {Kovtyukh}, {Usenko}, \&
  {Gorlova}}]{2013ApJ...762L...8T}
{Turner}, D.~G., {Kovtyukh}, V.~V., {Usenko}, I.~A., \& {Gorlova}, N.~I. 2013,
  \apjl, 762, L8

\bibitem[{{Turner} {et~al.}(2005){Turner}, {Savoy}, {Derrah}, {Abdel-Sabour
  Abdel-Latif}, \& {Berdnikov}}]{2005PASP..117..207T}
{Turner}, D.~G., {Savoy}, J., {Derrah}, J., {Abdel-Sabour Abdel-Latif}, M., \&
  {Berdnikov}, L.~N. 2005, \pasp, 117, 207

\bibitem[{{Usenko} {et~al.}(2005){Usenko}, {Miroshnichenko}, {Klochkova}, \&
  {Yushkin}}]{2005MNRAS.362.1219U}
{Usenko}, I.~A., {Miroshnichenko}, A.~S., {Klochkova}, V.~G., \& {Yushkin},
  M.~V. 2005, \mnras, 362, 1219

\bibitem[{{van Leeuwen}(2007)}]{2007ASSL..350.....V}
{van Leeuwen}, F., ed. 2007, Astrophysics and Space Science Library, Vol. 350,
  {Hipparcos, the New Reduction of the Raw Data}

\bibitem[{{van Leeuwen}(2013)}]{2013A&A...550L...3V}
{van Leeuwen}, F. 2013, \aap, 550, L3

\end{thebibliography}

\begin{appendix}
\section{Tables}
\subsection{Period-age relation for FO Cepheids on the first IS crossing}
\begin{table}[ht]
    \centering
    \begin{tabular}{lrrrr}
    \hline\hline
Metallicity  & B$_b$ & A$_b$ & B$_r$ & A$_r$  \\
\hline
Solar (0.014) & -0.890 & 8.267 & -0.867 & 8.422\\
LMC (0.006) & -0.944 & 8.264 & -0.923 & 8.432\\
SMC (0.002) & -0.993 & 8.265 & -0.974 & 8.446\\
\hline\hline
    \end{tabular}
    \caption{Period-age relations of the form $\log{(a\,\rm{[yr]})} = A + B \cdot \log{(P\,\rm{[d]})}$ 
    for FO Cepheids on the first IS crossing based on fit of predictions provided by A16. Subscripts $_b$ and $_r$ denote relations valid at the hot (blue) and cool (red) IS boundary, respectively.}
    \label{tab:period-age}
\end{table}

\subsection{Period-radius relations for FO Cepheids on the first IS crossing}
\begin{table}[ht]
    \centering
    \begin{tabular}{lrrrr}
    \hline\hline
Metallicity  & B$_b$ & A$_b$ & B$_r$ & A$_r$  \\
\hline
Solar (0.014) & 0.748 & 1.256 & 0.737 & 1.232\\
LMC (0.006) & 0.760 & 1.250 & 0.751 & 1.223\\
SMC (0.002) & 0.764 & 1.240 & 0.756 & 1.212\\
\hline\hline
    \end{tabular}
    \caption{Period-radius relations of the form $\log{(R/\rm{R}_\odot)} = A + B \cdot \log{(P\,\rm{[d]})}$ 
    for FO Cepheids on the first IS crossing based on fit of predictions provided by A16. Subscripts $_b$ and $_r$ denote relations valid at the hot (blue) and cool (red) IS boundary, respectively.}
    \label{tab:period-radius}
\end{table}

\end{appendix}

\end{document}